\documentclass[10pt,conference]{IEEEtran}
\IEEEoverridecommandlockouts
\usepackage[utf8]{inputenc}
\usepackage{graphicx} 
\usepackage{booktabs} 
\usepackage{amsmath}
\usepackage{bm}
\usepackage{gensymb}
\usepackage{physics}
\usepackage{mathtools, nccmath}
\usepackage{algorithm}
\usepackage{algpseudocode}
\usepackage{setspace}
\usepackage{amsfonts}
\usepackage{amssymb}
\usepackage{bbm}
\usepackage{cite}
\usepackage{lipsum}
\usepackage{authblk}
\usepackage{adjustbox}
\usepackage{comment}
\usepackage{soul}
\usepackage{todonotes}

\title{Fully Asynchronous  Unsourced Random Access over Fading Channels}
\author[*]{Mert Ozates\thanks{Mert Ozates' work is supported by the European Union’s (EU’s) Horizon Europe project 6G-SENSES under Grant 101139282. Gianluigi Liva's work is supported by the Federal Ministry of Education and Research of Germany in the programme of “Souverän. Digital. Vernetzt” joint Project 6G-RIC, project identification number: 16KISK026. Mohammad Kazemi’s work was funded by UK Research and Innovation (UKRI) under the UK government’s Horizon Europe funding guarantee [grant number 101103430]. Deniz Gündüz's work is supported by SNS JU project 6G-GOALS under the EU’s Horizon program (Grant Agreement No. 101139232).}}
\author[**]{Mohammad Kazemi}
\author[***]{Gianluigi Liva}
\author[**]{Deniz Gündüz}
\affil[*]{IHP - Leibniz Institute for High Performance Microelectronics, 15236 Frankfurt (Oder), Germany \protect\\ Email: oezates@ihp-microelectronics.com}
\affil[**]{Department of Electrical and Electronic Engineering, Imperial College London  \protect\\ Email:\{mohammad.kazemi, d.gunduz\}@imperial.ac.uk}
\affil[***]{Institute for Communications and Navigation, German Aerospace Center (DLR), Wessling, Germany \protect\\Email: gianluigi.liva@dlr.de }
\date{}

\begin{document}

\maketitle
\thispagestyle{empty}
\pagestyle{empty} 

\begin{abstract}
We examine unsourced random access in a fully asynchronous setup, where active users transmit their data without restriction on the start time over a fading channel. In the proposed scheme, the transmitted signal consists of a pilot sequence and a polar codeword, with the polar codeword distributed across the data part of the packet in an on-off pattern. The receiver uses a double sliding-window decoder, where the inner window employs iterative decoding with joint timing and pilot detection, channel estimation, single-user decoding, and successive interference cancellation to recover the message bits, while the outer window enhances interference cancellation. The numerical results indicate that the proposed scheme exhibits only a slight performance loss compared to the synchronous benchmark while being more applicable in practice.

\end{abstract}


\section{Introduction}

 The increase in connected devices makes massive machine-type communications (mMTC) an essential aspect of 6G communication systems. In mMTC, a massive number of devices sporadically try to communicate with a base station without coordination. To enable reliable communication in such a scenario, the unsourced random access (URA) paradigm \cite{survey} is proposed in \cite{polyanskiy}, in which users share a common codebook to transmit their data. This allows the system to operate regardless of the total number of devices; only the number of active devices matters. The receiver aims to produce a list of the transmitted messages up to a permutation, and the main performance criterion is per-user probability of error (PUPE). In \cite{polyanskiy}, a random-coding achievability bound is presented that characterizes the energy-per-bit for the Gaussian multiple-access channel (MAC) without any complexity constraints. It is also shown that conventional random access approaches (e.g., ALOHA) exhibit a significant gap from this bound, underscoring the need to develop energy-efficient schemes with low complexity. Many schemes have since been developed for URA over Gaussian MACs \cite{vem,comp4,schiavone,odma1,odma2}, as well as single-antenna \cite{nassaji} and multiple-input multiple-output (MIMO) fading channels \cite{orth,twc,odma4,zhang}. Among the many proposed methods, on-off division multiple access (ODMA) in \cite{odma1,odma2,odma4} stands out for its high energy efficiency, scalability, and low complexity, thanks to its flexible sparsity. In ODMA, the transmitted codeword is sparsely distributed across the transmission frame according to a predetermined pattern, while a pilot sequence can be appended to enable channel estimation at the receiver when applied to fading channels.


The studies mentioned above assume perfect synchronization, i.e., there is an aligned transmission frame, which is hard to achieve in practice. As a result, many schemes are developed for asynchronous transmission in the literature, e.g., for satellite communications \cite{gallinaro,herrero}. Synchronous transmission is particularly challenging in URA systems, as synchronizing a large number of devices with limited power and computational capabilities is difficult. 
The ALOHA protocol and its variants provide solutions for asynchronous setup and are also applicable to the URA scenario. However, they are inefficient in a URA setup due to their limited interference-resolution capabilities \cite{polyanskiy}, necessitating URA-targeted solutions.
In earlier asynchronous URA works, a limited delay among user transmissions is assumed, and orthogonal frequency division multiplexing (OFDM) is employed to tackle the problem in the frequency domain, using a cyclic prefix longer than the maximum delay. On the other hand, a fully asynchronous setup in which the users transmit their data without any restriction on the starting time is studied in \cite{karami} and \cite{ozatesfully}. In \cite{karami}, a preamble is utilized to estimate the transmission starting times, and the payload is transmitted after permutation and scrambling. An ODMA-based preamble-free scheme employing polar codes is proposed in \cite{ozatesfully}, where the transmission start times are estimated from ODMA patterns, and a double sliding-window decoder is used at the receiver. However, both works address the problem of Gaussian MACs, and, with reference to the URA setup of \cite{polyanskiy}, we are not aware of any work that addresses fully asynchronous massive access on fading channels. 



To fill this gap, in this paper, we study fully asynchronous URA explored in \cite{karami} and \cite{ozatesfully} over a fading channel, and propose an ODMA-based scheme that exploits pilots for both timing detection and channel estimation. We assume that the users divide their packets into two parts. In the first part (pilot part), a pilot sequence selected from a common codebook based on a part of the message bits is transmitted. The remaining bits are encoded by a polar code, modulated, and distributed to the data part in an ODMA manner. At the receiver side, we utilize a sliding-window decoding algorithm with two nested windows, similarly to \cite{ozatesfully}. In the inner window, decoding operations of joint pilot and starting time detection, channel estimation, and single-user decoding, followed by successive interference cancellation (SIC), are iteratively performed to recover the transmitted messages, with the re-estimation of the channel coefficients at the end of each iteration, while the outer window improves interference cancellation. Since no prior work has studied fully asynchronous URA in the fading setup, we compare the performance of our proposed scheme with that of its synchronous counterpart. Numerical examples illustrate that, although the proposed scheme is fully practical and transmissions can be performed without any restriction on their starting time, its performance gap with its synchronous counterpart is less than 1 dB when the number of packet arrivals per packet duration is less than 100. 



The rest of the paper is organized as follows. We introduce the system model in Section \ref{system} and the proposed scheme in Section \ref{proposed}. A set of numerical results are presented in Section \ref{results}, and the paper is concluded in Section \ref{conclusion}.

\section{System Model} \label{system}

We consider a URA scenario in which $K_a$ active users transmit $B$ bits of data to a common base station (BS) equipped with a single antenna over a quasi-static fading channel, in a fully asynchronous manner. Namely, there is no frame alignment, and a user transmits its data whenever it is generated, i.e., there is no restriction on the transmission starting time. The received signal at the $j$-th time instance can be written as

\begin{equation}
    \mathbf{y}_j =  \sum\limits_{i \in \mathcal{K}_j} \mathbf{x}_i(j-\delta_i) h_i + \mathbf{z}_j,
\end{equation}

\noindent where $\delta_i$ denotes the starting time of the transmission of the $i$-th user, $\mathcal{K}_j$ is the set of active users at the discrete time instance $j$, $h_i$ is the complex fading coefficient of the $i$-th user, and $\mathbf{z}_j$ is the complex additive  white Gaussian noise (AWGN) with zero-mean and variance $\sigma^2$. Moreover, $\mathbf{x}_i$ denotes the transmitted signal of the $i$-th user with a packet length of $n$. We assume that $h_i$ is a complex Gaussian random variable with zero mean and unit variance, i.e., $h_i \sim \mathcal{CN}(0,1)$, its value remains constant during one active period of the user and changes in an independent and identically distributed (i.i.d.) fashion in the next activation period. In our proposed scheme, first $n_p$ symbols of a packet are allocated for pilot transmission, while $n_d$ of the remaining $n-n_p$ symbols are used for data transmission, i.e., the codeword symbols are placed in these indices (active indices), and the remaining $ n - (n_p + n_d) $ instances are idle. The transmitted signal satisfies the power constraint of $\norm {\mathbf{x}_i}^2 \leq  n_pP_p + n_dP_d $, where $P_p$ and $P_d$ are the average symbol power of the pilot and data part, respectively. The energy per bit of the system can be calculated as

\begin{equation}
    \frac{E_b}{N_0} =  \frac{n_dP_d + n_pP_p}{B\sigma^2}.
\end{equation}

We assume, without loss of generality, that transmissions begin after time $t = 0$. We observe the channel output over a time interval of $[0, T + n]$, where $T \gg n$, and compute the PUPE as 

\begin{equation}
P_e =\frac{1}{K_{a,T}} \sum\limits_{i=1}^{K_{a,T}} \Pr(\mathbf{m}_i \notin \mathcal{L}),
\end{equation}

\noindent where $\mathcal{L}$ is an unordered list of transmitted messages extracted at the receiver, $K_{a,T}$ is the number of active users in $[0, T]$, i.e., the number of users initiating a transmission in $[0, T]$, and $\mathbf{m}_i$ denotes the message from the $i$-th user. Note that the average number of packet arrivals in a packet duration $n$ is denoted by $K_a$, which is assumed to be known to the receiver (see \cite{twc} for how to estimate $K_a$ when it is unknown). Finally, the goal is to minimize the energy per bit required to achieve a target PUPE of $\epsilon$.

\section{Proposed Scheme} \label{proposed}

\subsection{Encoding}

\begin{figure}
    \centering
    \hspace*{-5mm}
     \includegraphics[scale = 0.35]{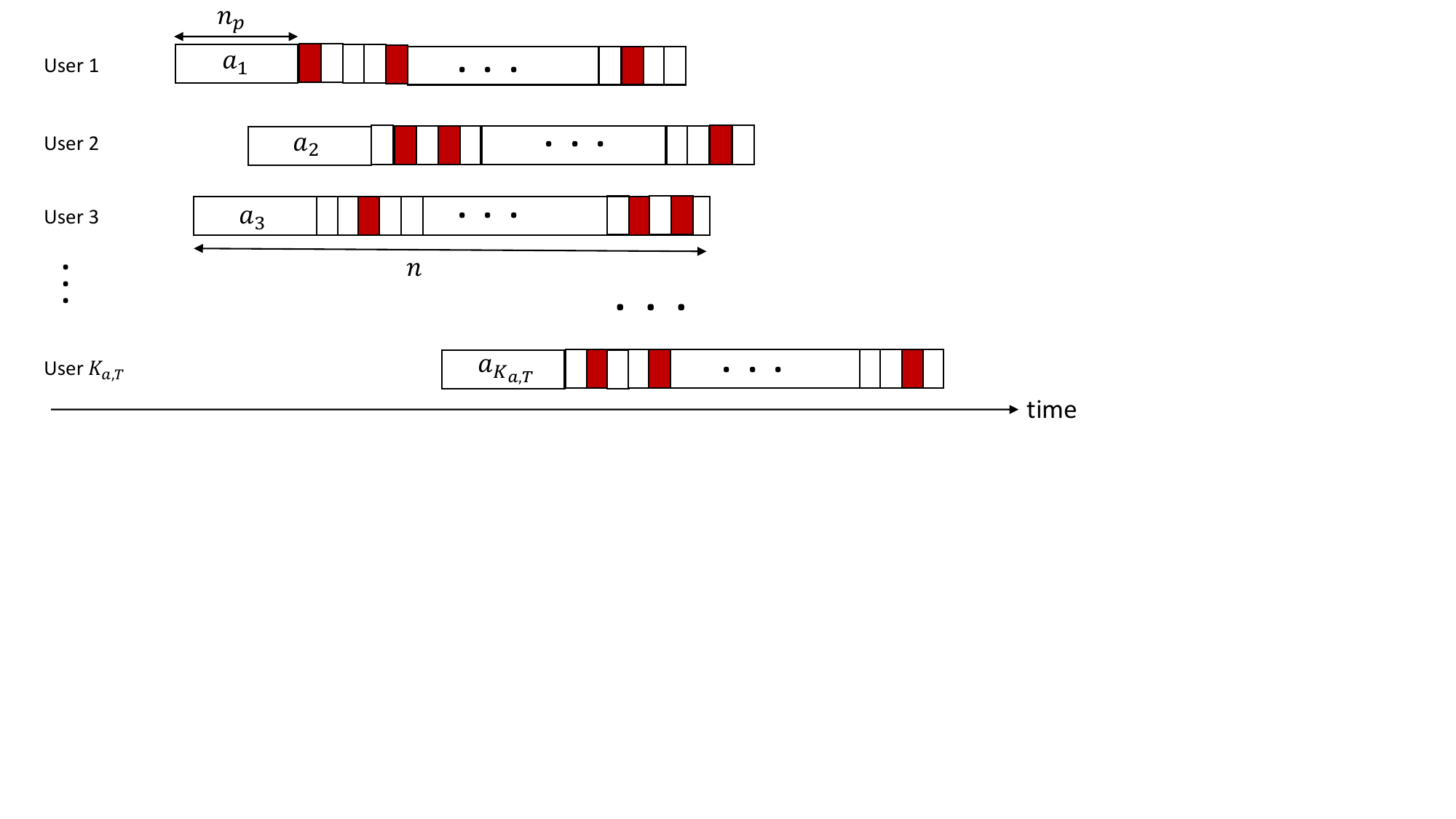}
     \vspace*{-35mm}
    \caption{The proposed fully asynchronous transmission structure for fading channels. Transmitted packets consist of a pilot and a data part. Colored boxes ($n_d$ per packet) in the data part indicate the used symbols.}
    \label{encoder}
\end{figure}

In our proposed scheme, the message sequence is divided into two parts. Each user picks a pilot sequence from a common pilot codebook $\mathbf{A} \in \mathbb{C}^{n_p \times N}$ to be transmitted in the first $n_p$ symbol instances of the packet, selected based on the first $B_p$ bits of its message, where $N= 2^{B_p}$. The rest of the bits are encoded by a $(n_c, B-B_p + r)$ polar code and modulated by binary phase shift keying (BPSK), where $n_c$ is the polar code length and $r$ is the number of cyclic redundancy check (CRC) bits. The encoded and modulated bits are then randomly placed in the last $n-n_p$ symbol instances based on a transmission pattern that is picked randomly and independently from a common pattern matrix $\mathbf{P} \in \{0,1\} ^{n-n_p \times N}$. Note that the transmission pattern is also dictated by the first $B_p$ bits. Each column of $\mathbf{P}$ contains $n_d$ non-zero elements ($n_d = n_c$ for BPSK) determining the placement of transmitted symbols, while the remaining elements represent the idle indices. The proposed fully asynchronous transmission structure, with transmitted signals consisting of a pilot and a data part, is shown in Fig. \ref{encoder}. 

\subsection{Decoding Procedure}

\begin{figure}
    \centering
     \includegraphics[scale = 0.43]{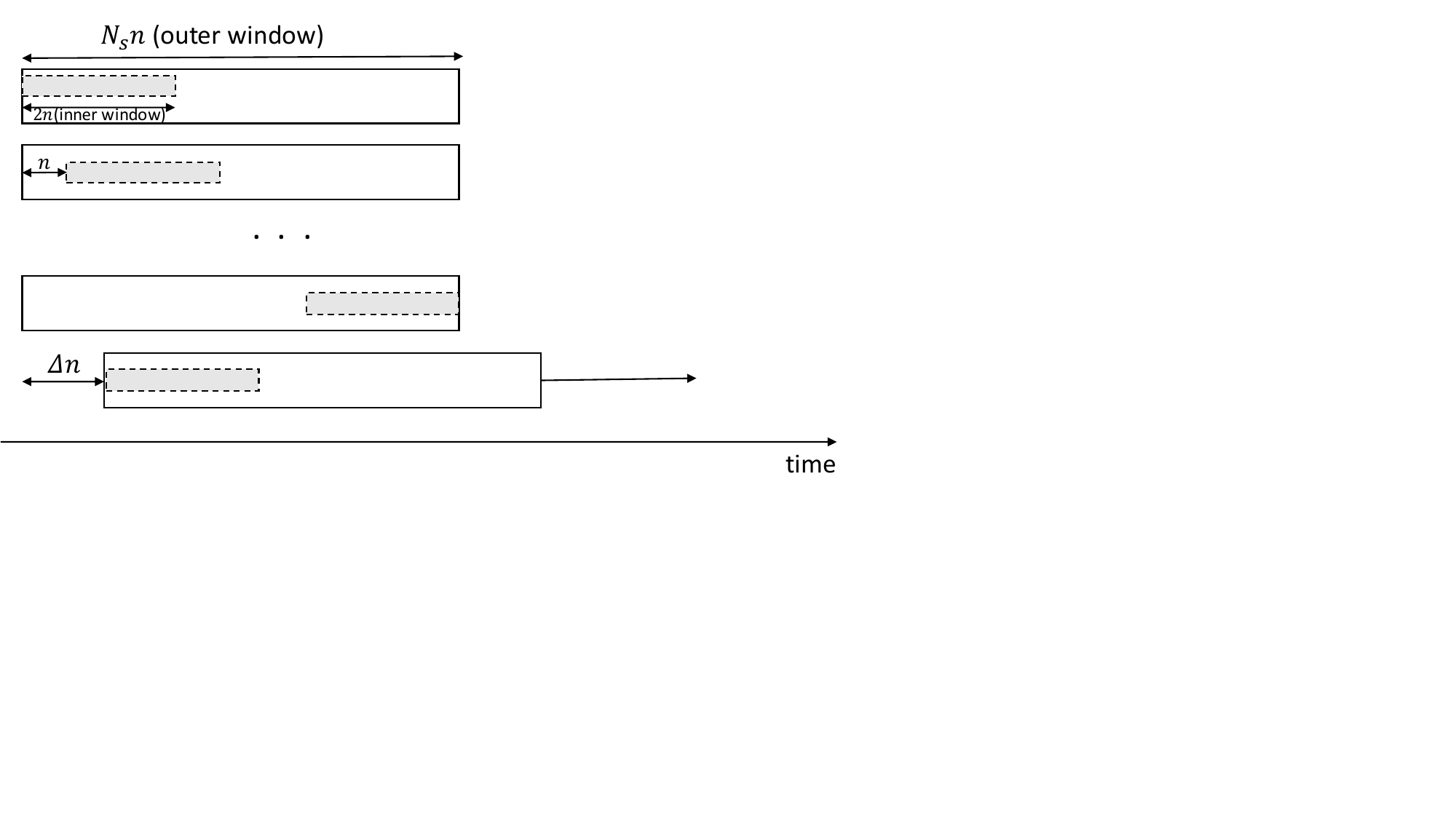}
    \vspace*{-40mm}
    \caption{Nested operation of sliding window decoding.}
    \label{figslide}
\end{figure}

At the receiver side, we utilize a sliding-window based decoding algorithm consisting of two nested windows, similarly to \cite{ozatesfully}. The outer sliding window is of $N_s$ packet lengths, while the inner sliding window is a window of two packet lengths. In the inner sliding window, the aim is to decode the messages of the active users, starting from the first up to the $n$-th time instance of the window. For this purpose, the transmission start times and pilot sequences of the users are first jointly estimated. Then, the channel coefficients are estimated separately for each detected starting time and pilot pair, followed by single-user decoding and SIC. For SIC, the channel coefficients are jointly re-estimated to improve the decoding performance. After the inner-window decoding operations are completed, the inner window is shifted by one packet length, and the procedure is repeated until the inner window reaches the end of the outer window, which marks the end of one outer iteration. Then the inner window is moved back to the beginning of the outer window, and the same procedure is applied. This enables decoding more users with reduced interference, which is achievable with another window. After completion of the outer iterations, the outer window is shifted by $\Delta$ packet lengths, and the same procedure is applied. The operation of two nested sliding windows is illustrated in Fig. \ref{figslide}. In the following, we explain the operations in the inner decoding window in detail.

\subsubsection{Joint Starting Time and Pilot Estimation}

We first jointly detect the transmission starting times and selected pilot sequences using the pilot codebook. To do that, we correlate each sequence in our pilot codebook with the received signal portions of length $n_p$, where the signal portions are the received signal blocks starting in the first half of the inner window. In other words, we test each pilot at each time instance $b = 1, 2, \dots, n$ by calculating the correlation between the pilot sequences and the received signal portions as follows:

\begin{equation}
    \mathbf{R}_{b} =  \mathbf{A}^H  \mathbf{y}_b, \quad b=1, \dots, n, 
    \label{eqtime}
\end{equation}

\noindent where $\mathbf{R}_b \in \mathbb{C}^{N \times 1}$, $\mathbf{y}_b$ is the received signal portion of length-$n_p$ in the inner sliding window $\mathbf{y}_s$, beginning at time instance $b$. For simplicity, we omit the iteration index here and in the rest of the subsection; namely, the equations hold for the first iteration and are applied to the residual signals in subsequent iterations. Then, for each time instance, the pilot sequence with the largest absolute correlation is selected as the survivor pilot. After the calculation is performed for each time instance, $K_s = K_a + u$ of the $n$ survival pilots are taken as the detected pilots, and their corresponding time instances become the detected transmission starting times, where $u$ is a margin on the average active user load, accounting for the fluctuations in the number of active users starting their transmissions within each block of length $n$. Note that the transmission patterns are inherently identified in this step, as they share the same index with the pilots, since they are both determined by the first $B_p$ bits.

\subsubsection{Channel Estimation}

Given the starting time instances and pilot sequences, the channel coefficients of the users are estimated using a linear minimum mean-square error (MMSE) solution as follows.

\begin{equation}
    \hat{\mathbf{h}} = (\mathbf{\hat{A}}^H\mathbf{\hat{A}} + N_0\mathbf{I}_{K_s})^{-1} \mathbf{\hat{A}}^H \mathbf{y}_s,
    \label{eqest}
\end{equation}

\noindent where $\hat{\mathbf{h}} \in \mathbb{C}^{K_s \times 1}$ and $\mathbf{\hat{A}} \in \mathbb{C}^{2n \times K_s} $ is the set of estimated pilot sequences with delays. The $i$-th column of $\mathbf{\hat{A}}$ is formed by placing the estimated pilot sequence of the $i$-th user starting its estimated transmission time, while the other elements remain idle.


\subsubsection{Single-User Decoding}

Following channel estimation, for each transmission starting time, channel, and pattern estimate triplet, the log-likelihood ratios (LLRs) corresponding to the transmitted symbols are calculated from the received signal points at the active indices, treating interference as noise (TIN). Note that given the transmission patterns, the received signal at the active indices becomes the output of a single-user quasi-static fading channel, assuming that the multiuser interference is Gaussian. The LLRs are then fed to a polar decoder employing successive cancellation list decoding (SCLD), and the decoded sequence is added to a temporary output list $\hat{\mathcal{D}}$ if it satisfies the CRC check.

\subsubsection{SIC}

To subtract the effect of the successfully decoded messages, they are first re-encoded and modulated. Then, instead of using the initially estimated channel coefficients, the channel coefficients of the successfully decoded users are jointly re-estimated using the decoded sequences along with the corresponding pilot sequences as

\begin{equation}
    \mathbf{\hat{h}}_{\text{SIC}} = (\mathbf{\hat{X}}_{\hat{\mathcal{D}}}^H\mathbf{\hat{X}}_{\hat{\mathcal{D}}} + \sigma^2 \mathbf{I}_{\abs{\mathcal{\hat{D}}}} )^{-1} \mathbf{\hat{X}}_{\hat{\mathcal{D}}}^H \mathbf{y}_s,
    \label{eqreest}
\end{equation}

\noindent where $\mathbf{\hat{X}}_{\hat{\mathcal{D}}} \in \mathbb{C}^{2n \times {\abs{\mathcal{\hat{D}}}} }$ is the set of reconstructed transmitted symbols. Each column of $\mathbf{\hat{X}}_{\hat{\mathcal{D}}}$ is formed by placing the reconstructed transmitted signal of the corresponding user, i.e., its estimated pilot, followed by the re-encoded and modulated signal distributed by ODMA to the data part, starting at its detected transmission starting time. Then, SIC is performed using the re-estimated channel coefficients. The decoding iterations continue until no new message can be decoded, or a maximum number of iterations $n_{\text{max}}$ is reached.

\begin{figure}
    \centering
    \includegraphics[trim={0 13cm 0 1cm}, clip, width=1.5\linewidth]{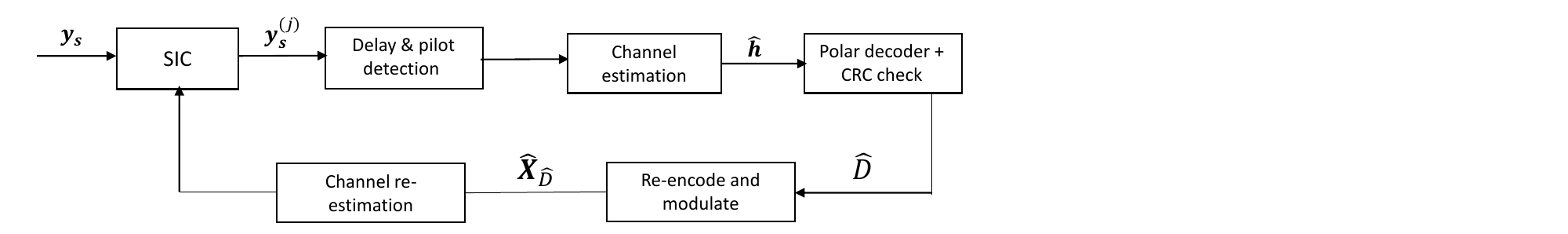}
    \caption{The decoding process of the inner sliding window.}
    \label{decoder}
\end{figure}

After the iterations of the inner sliding window are completed, it is shifted by one packet length of $n$ channel symbols, and the same procedure is repeated until the end of the outer window (i.e., $N_s -1$ times), corresponding to one outer iteration. The pseudo-code for the decoding operation of the inner sliding window is provided in Algorithm \ref{alg_inner}, and illustrated in Fig. \ref{decoder}.

 \begin{algorithm}[t]
\caption{Decoding process of the inner sliding window.}\label{alg_inner}
\begin{algorithmic}[1]

\State \textbf{Input}: $\mathbf{y}_s$,  $\mathbf{P}$, $\mathbf{A}$, $u$, $n_{\text{max}}$
\State \textbf{Iterative decoding}:
\State Set $\mathcal{\hat{D}} = \emptyset$
\For{$j=1, \ldots, n_{\text{max}} $}
\State \textbf{Joint starting time and pilot estimation}:

\For{$b=1, \ldots, n $}

\State $\mathbf{y}_b$ = $\mathbf{y}_s[b:b+n_p-1]$.
\State $\mathbf{R}_{b} =  \mathbf{A}^H  \mathbf{y}_b$
\EndFor
\State Take indices of highest $K_a + u$ elements among $\mathbf{R}_{1}, \dots ,\mathbf{R}_{n}$ as starting time estimates.
\State \textbf{Channel estimation}:
\State Estimate channel coefficients by (\ref{eqest})
\State \textbf{Single-user decoding and SIC}:
\State Set $d = 0$
\For{$i=1,\ldots, {K}_a + u$}
\State LLR extraction by TIN
\State Polar decoding $\rightarrow$ $\mathbf{\hat{m}}_i$
\If {CRC check is successful:}
\State Add $\mathbf{\hat{m}}_i$ to output list
\State d = d + 1
\EndIf
\EndFor
\State Re-estimate channel coefficients by (\ref{eqreest})
\State Perform SIC
\If {$  d = 0$}
\State Terminate
\EndIf
\EndFor


\State $\textbf{Output}$: Intermediate list of decoded messages $ \mathcal{\hat{D}} $

\end{algorithmic}
\end{algorithm}

\section{Numerical Results} \label{results}
We assess the performance of the proposed scheme through Monte Carlo simulations. We take the packet length as $n = 10000$, the message length $B = 100$, and set the target PUPE to $\epsilon = 0.1$. We utilize a randomly generated binary matrix of size $(n-n_p) \times N$ with a column weight of $n_d$, where $N = 16$ ($B_p = 4$) for transmission patterns, and set the pilot length to $n_p = 3000$. The elements of the pilot sequences are drawn from a standard Gaussian distribution, and each sequence is normalized to have a norm of $\sqrt{n_pP_p}$ to satisfy the power constraint. Note that pilot/pattern collisions may cause a decoding failure only if multiple users start the transmission at the same time instance, which has a very small probability in the asynchronous case. Hence, a small number of pilots/patterns can be used. We employ 5G polar codes, and set the polar code length to 512, the CRC length to 16, the SCLD list size to 32, $n_{\text{max}} = 50$, and $n_{\text{out}} = 10$, that is the maximum number of outer iterations. The outer sliding-window length is taken as $N_s = 5$ packet lengths, and it is shifted by one packet length after completing the outer iterations (i.e., $\Delta  = 1$). Note that the inner sliding window length is fixed to two packet lengths. To improve decoding performance, we also subtract the effect of successfully decoded sequences within the inner iterations. Since our work is the first to study fully asynchronous URA over fading channels, there are no schemes in the literature for direct comparison. Hence, we take the perfect synchronization scenario proposed in \cite{odma2} as a benchmark and compare the performance of the proposed scheme with that of the synchronous case.

\begin{figure} 
    \centering
    \includegraphics[trim={0 0 0 .7cm}, clip, width=1\linewidth]{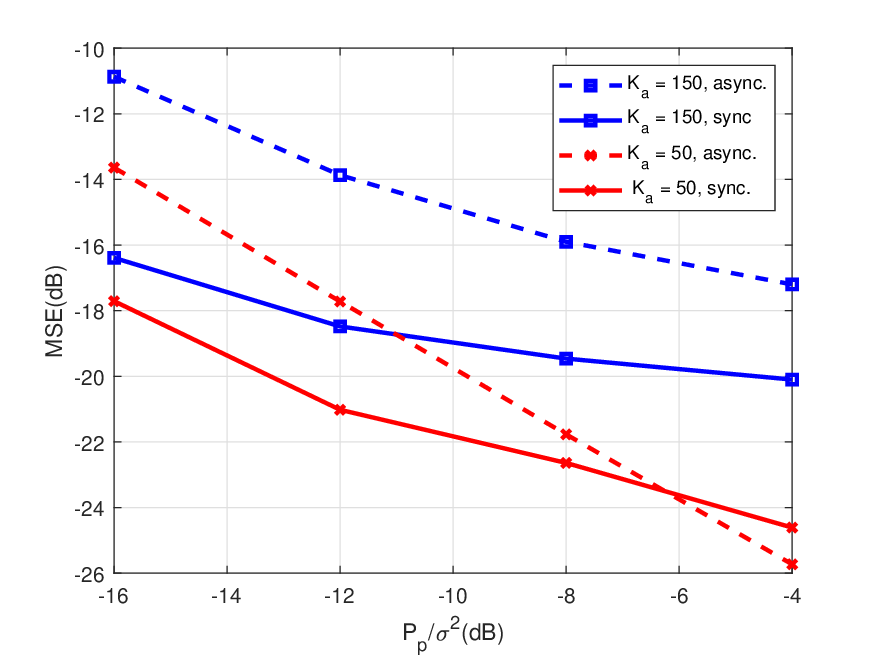}
    \caption{Comparison of channel estimation MSE of synchronous and asynchronous case for $K_a  = 50,150$.}
    \label{figmse}
\end{figure}

We first compare the channel estimation performance of the synchronous and asynchronous cases for $K_a = 50,150$ and $P_d/\sigma^2 = 10$ dB by calculating the mean square error (MSE) between the actual and estimated channel coefficients as follows:

\begin{equation}
    \text{MSE} = \frac{\sum\limits_{i=1}^{K_a} \norm{\hat{h}_i - h_i } ^2 }{K_a}.
\end{equation}

The channel estimation performance with respect to the pilot transmission power, for a fixed data transmission power, is presented in Fig. \ref{figmse}. It can be seen that the channel estimation MSE in the synchronous case is lower than in the asynchronous case by up to 4 dB for $K_a = 50$ in the low-power regime; however, the gap decreases with increasing pilot power. Note that the asynchronous case can even be better when the pilot power is sufficiently high. A similar behavior is observed for $K_a = 150$; that is, the MSE in the synchronous case is reduced by 3-5 dB, and the gap decreases as the pilot power increases. Note that there is no data interference in the synchronous case because the user transmissions are aligned, whereas in the asynchronous case, data interference acts as additional noise. However, its effect decreases as pilot power increases.



\begin{figure} 
    \centering
    \includegraphics[width=1\linewidth]{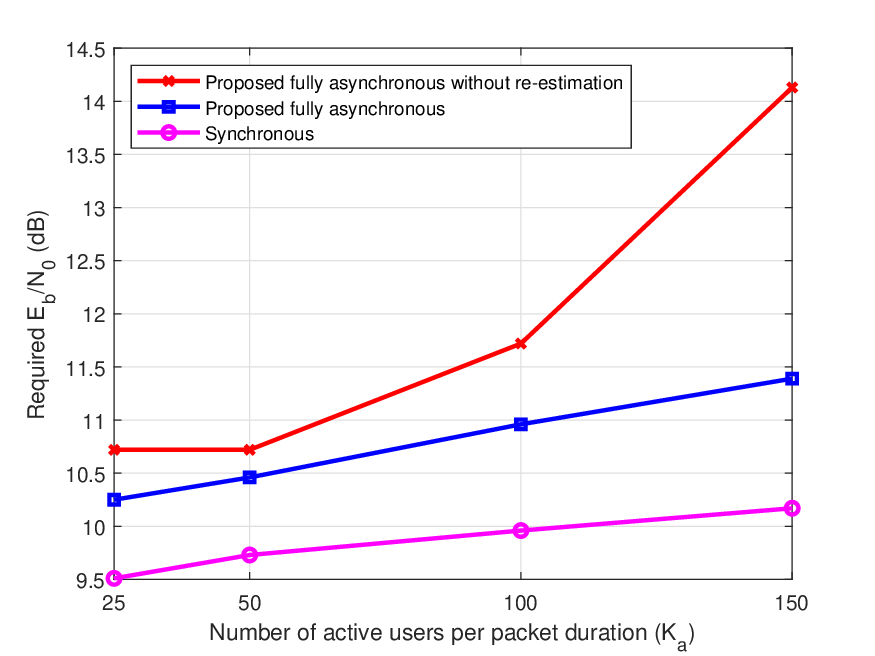}
    \caption{Required energy-per-bit versus the number of active users per packet duration for $P_e \leq 0.1.$}
    \label{figeff}
\end{figure}

We assess the energy efficiency of the proposed scheme by calculating the required energy per bit and comparing it with that of the synchronous benchmark. In the synchronous case, the number of pilot bits is set to $B_p = 13$ to avoid collisions, resulting in a slightly lower coding rate. The results in Fig. \ref{figeff} demonstrate that the performance difference of the proposed scheme and the synchronous case is at most approximately 1 dB. Our extensive simulations show that, in the operating pilot power regime to achieve a target PUPE of 0.1, the channel estimation performance in the asynchronous scenario is inferior to that in the synchronous scenario, resulting in inferior system performance. Another reason is the difference in coding rate, as lower coding rates improve system performance. In addition, re-estimating the channel coefficients improves the performance up to 2.5 dB.



\section{Conclusions} \label{conclusion}

We proposed an ODMA-based transmission scheme for fully asynchronous URA over a fading channel, employing pilots for timing detection and channel estimation, and polar coding for data transmission. At the receiver side, we employed a sliding-window decoding algorithm with two nested windows, which improves decoding performance by enhancing interference cancellation. The numerical results show that the proposed scheme offers a close performance to the synchronous scenario when the number of active users is not too large.



\begin{thebibliography}{}

\bibitem{survey}M. Ozates, M. J. Ahmadi, M. Kazemi, D. Gündüz and T. M. Duman, "Unsourced random access: A comprehensive survey," {\it IEEE Commun. Surveys Tuts.}, early access, 2025.


\bibitem{polyanskiy} Y. Polyanskiy, \textquotedblleft{}A perspective on massive random-access,\textquotedblright{} in \textit{Proc. IEEE Int. Symp. Inf. Theory (ISIT)}, Aachen, Germany, June 2017, pp. 2523-2527.

\bibitem{vem} A. Vem, K. R. Narayanan, J. -F. Chamberland and J. Cheng, \textquotedblleft{}A user-independent successive interference cancellation based coding scheme for the unsourced random access Gaussian channel," \textit{IEEE Trans. Commun.}, vol. 67, no. 12, pp. 8258-8272, Dec. 2019.

\bibitem{comp4} E. Nassaji and D. Truhachev, “Dynamic compressed sensing approach for unsourced random access,” {\it IEEE Commun. Lett.}, vol. 28, no. 7, pp. 1644-1648, July 2024.

\bibitem{schiavone} R. Schiavone, G. Liva and R. Garello, “Design and performance of enhanced spread spectrum aloha for unsourced multiple access,” {\it IEEE Commun. Lett.}, vol. 28, no. 8, pp. 1790-1794, Aug. 2024.


\bibitem{odma1} J. Yan, G. Song, Y. Li and J. Wang, \textquotedblleft{}ODMA transmission and joint pattern and data recovery for unsourced multiple access," \textit{IEEE Wireless Commun. Lett.}, vol. 12, no. 7, pp. 1224-1228, July 2023.

\bibitem{odma2} M. Ozates, M. Kazemi and T. M. Duman, \textquotedblleft{}Unsourced random access using ODMA and polar codes," \textit{IEEE Wireless Commun. Lett.}, vol. 13, no. 4, pp. 1044-1047, Apr. 2024.

\bibitem{nassaji} E. Nassaji, M. Bashir and D. Truhachev, \textquotedblleft{}Unsourced random access over fading channels via data repetition, permutation, and scrambling," \textit{IEEE Trans. Commun.}, vol. 70, no. 2, pp. 1029-1042, Feb. 2022.


\bibitem{orth} M. J. Ahmadi, M. Kazemi and T. M. Duman, \textquotedblleft{}Unsourced random access using multiple stages of orthogonal pilots: MIMO and single-antenna structures," \textit{IEEE Trans. Wireless Commun.}, vol. 23, no. 2, pp. 1343-1355, Feb. 2024.

\bibitem{twc} M. Ozates, M. Kazemi and T. M. Duman, \textquotedblleft{}A slotted pilot-based unsourced random access scheme with a multiple-antenna receiver," \textit{IEEE Trans. Wireless Commun.}, vol. 23, no. 4, pp. 3437-3449, Apr. 2024.

\bibitem{odma4} M. Ozates, M. Kazemi and T. M. Duman, “An ODMA-based unsourced random access scheme with a multiple antenna receiver,” in {\it Proc. IEEE Global Commun. Conf. (GLOBECOM)}, Cape Town, South Africa, 2024, pp. 1857-1862.

\bibitem{zhang} Z. Zhang, J. Dang and Z. Zhang, “Unsourced random access under quasi-static fading channel: A less-is-more strategy," \textit{IEEE Trans. Wireless Commun.}, early access.

\bibitem{gallinaro} G. Gallinaro, N. Alagha, R. De Gaudenzi, K. Kansanen, R. Müller and P. Salvo Rossi, \textquotedblleft{}ME-SSA: An advanced random access for the satellite return channel," in {\it Proc. IEEE Int. Conf. Commun.(ICC)}, London, UK, 2015, pp. 856-861.

\bibitem{herrero} O. Del Rio Herrero and R. De Gaudenzi, “A high efficiency scheme
for quasi-real-time satellite mobile messaging systems,” in {\it Proc. Int.
Workshop Signal Processing for Space Commun.}, Oct. 2008.

\bibitem{kowshik} S. S. Kowshik, K. Andreev, A. Frolov and Y. Polyanskiy, “Short-packet low-power coded access for massive multiple access,” in {\it Proc. 53rd Asilomar Conference on Signals, Systems, and Computers}, CA, USA, Sep. 2019, pp. 827-832.

\bibitem{mimoasync1} W. Wang, J. You, S. Liang, W. Han and B. Bai, “Slotted concatenated coding scheme for asynchronous uplink unsourced random access with a massive MIMO receiver,” in \textit{Proc. IEEE 33rd Annual Int. Symp. Personal, Indoor and Mobile Radio Commun. (PIMRC)}, Kyoto, Japan, 2022, pp. 246-252.

\bibitem{mimoasync2}  T. Li et al., “Asynchronous MIMO-OFDM massive unsourced random access with codeword collisions," {\it IEEE Trans. Wireless Commun.}, vol. 24, no. 1, pp. 84-100, Jan. 2025.

\bibitem{karami} A. Karami, I. Pazouki, R. Soltani and D. Truhachev, \textquotedblleft{}An approach to asynchronous unsourced random access (URA),\textquotedblright{} in \textit{Proc. IEEE Wireless Commun. Netw. Conf. (WCNC)}, Dubai, United Arab Emirates, 2024, pp. 01-06. 

\bibitem{ozatesfully} M. Ozates, M. Kazemi, G. Liva and D. Gündüz, “A fully asynchronous unsourced random access scheme," in {\it Proc. IEEE 26th Int. Workshop Signal Process. Artif. Intell. Wireless Commun. (SPAWC)}, Surrey, United Kingdom, 2025, pp. 1-5.



\end{thebibliography}
\end{document}